**Strained 2D TMD lateral heterojunctions via grayscale thermal-Scanning Probe Lithography**

*Giorgio Zambito[1], Giulio Ferrando[1], Matteo Barelli[1], Michele Ceccardi[1], Federico Caglieris[2], Daniele Marre[1], Francesco Bisio[2], Francesco Buatier de Mongeot[1]*, Maria Caterina Giordano[1]*

Giorgio Zambito[1], Giulio Ferrando[1], Matteo Barelli[1], Michele Ceccardi[1], Daniele Marre[1], Francesco Buatier de Mongeot[1]*, Maria Caterina Giordano[1]

Dipartimento di Fisica, Università di Genova, Via Dodecaneso 33, 16146, Genova, Italy

E-mail: buatier@fisica.unige.it

Federico Caglieris[2], Francesco Bisio[2]

CNR-SPIN, Corso Perrone, 24, 16152, Genova, Italy

Funding:

M.C.G. acknowledges support under the National Recovery and Resilience Plan (NRRP) Call for tender n. 1409 PRIN 2022 PNRR by Ministero dell'Università e della Ricerca (MUR), funded by the European Union – NextGenerationEU, project '2D-PLANET' (code P2022JWKAC), and Call for tender n. 104 PRIN 2022, project 'design' (code 2022EE8KH9).
M.C.G. acknowledges support by Università degli Studi di Genova "Progetto Curiosity Driven 2021", funded by the European Union – NextGenerationEU.
F.B.d.M., D.M. and M.C.G. acknowledge support of MUR, Project funded under the National Recovery and Resilience Plan (NRRP), Mission 4 Component 2 Investment 1.3 - Call for tender No. 1561 of 11.10.2022; funded by the European Union – NextGenerationEU • Award

Number: Project code PE0000021, Concession Decree No. 1561 of 11.10.2022 adopted by MUR.
F.B.d.M. acknowledges support of MUR, project PRIN2022 “2022FWB2HE”, and of MASE (Ministero dell’Ambiente e della Sicurezza Energetica) in the framework of the Operating Agreements with ENEA for Research on the Electric System.
F. B. acknowledges financial support under the National Recovery and Resilience Plan (NRRP), Mission 4, Component 2, Investment 1.1, Call for tender no. 104 PRIN 2022 published on 2.2.2022 by the Italian Ministry of University and Research (MUR), funded by the European Union – NextGenerationEU –Project Title ERACLITO – CUP B53D2300904006 and Call for tender no. 1409 PRIN PNRR 2022 published on 14.9.2022 by the Italian Ministry of University and Research (MUR), funded by the European Union – NextGenerationEU – Project Title ARCO –CUP B53D23027690001.

## Abstract

Nanoscale tailoring of the optoelectronic response of 2D Transition Metal Dichalcogenides semiconductor layers (TMDs) has been achieved thanks to a novel strain engineering approach based on the grayscale thermal-Scanning Probe Lithography (t-SPL). This method allows the maskless nanofabrication of locally strained 2D $MoS_2$-Au lateral heterojunction nanoarrays that are characterized by asymmetric electrical behavior.

2D $MoS_2$ layers are conformally transferred onto grayscale t-SPL templates characterized by periodic nanoarrays of deterministic faceted nanoridges. This peculiar morphology induces asymmetric and uniaxial strain accumulation in the 2D TMD material allowing to tailor their electrical work-function at the nanoscale level, as demonstrated by Kelvin Probe Force Microscopy (KPFM). The modulation of the electronic response has been exploited to develop periodic nanoarrays of lateral heterojunctions endowed with asymmetric electrical response by simple maskless deposition of Au nanocontacts onto the strained 2D TMD layers. The locally strained Au-$MoS_2$ layers show asymmetric lateral heterojunctions with engineered carrier extraction functionalities, thus representing a promising platform in view of tunable ultrathin nanoelectronic, nanophotonic and sensing applications.

## 1. Introduction

Transition Metal Dichalcogenides (TMDs) recently gathered increasing interest as promising two-dimensional (2D) materials for next generation ultrathin devices in various fields ranging from optoelectronics and biosensing to quantum technologies.[1–6] Thanks to their atomic layered structure and tunable electronic band structure, they have been exploited as building blocks for van der Waals heterostructures and represent optimal candidates for ultrathin nanoelectronics and nanophotonics applications [7–15]. A challenging aspect in this field deals with the fabrication of lateral heterojunctions between 2D materials that show different electronic properties, in order to engineer at will the optoelectronic response of ultra-thin devices[16–21]. In parallel, the exceptional mechanical resilience[22,23] of 2D materials offers the unique opportunity to tailor their optoelectronic response via strain engineering[24–28]. Strain dependent optoelectronic and photoemission properties have been so far observed by exploiting different approaches based either onto rough and nanopatterned materials[24,29–31], or onto large-area bending of the substrate[32–35]. Strain induced photonic effects have been recently demonstrated over large-area in self-organized nanogrooved templates[36,37]. However, these nanoarrays are characterized by a broad distribution of size, posing an issue in terms of the local control of the nanostructures morphology and order for the accurate tailoring of the strain-induced optoelectronic effects. Moreover, the optoelectronic properties of strained 2D semiconductors have been generally investigated via Raman and photoluminescence micro-spectroscopy with diffraction-limited spatial resolution at the micrometric scale [29,38]. Few pioneering experiments have been performed at the nanoscale by AFM probe-assisted bending with challenging detection that can be achieved either via complex nanomechanical devices or by diffraction-limited micro-spectroscopy[23,39,40].

The possibility to reshape fragile 2D layers at the nanoscale in a non-invasive way has been recently achieved by thermal- Scanning Probe Lithography. This technique allows local and controlled heating of the sample surface with a sharp nanotip and has been employed to perform thermochemical lithography, thermomechanical nanocutting, or straining of fragile 2D layers, showing promising results in photonics[41–45].

In this work, we demonstrate the active tailoring of the optoelectronic properties of 2D TMD semiconductor layers at the local scale with the aim of engineering 2D lateral junctions. To achieve this goal, we develop deterministic grayscale periodic templates with faceted morphology by a novel thermal-Scanning Probe Lithography approach, and we employ these platforms for the conformal transfer of 2D $MoS_2$ layers. Using Kelvin Probe Force Microscopy (KPFM) imaging of the periodically bent 2D layers we detected an asymmetric periodic modulation of the electrical surface potential induced by strain, with spatial resolution at the nanoscale. We exploit such strain induced modulation to engineer periodic lateral Au-$MoS_2$ heterojunctions with asymmetric response by simple maskless deposition of Au nanocontact nanoarrays. The KPFM imaging of these ultrathin hybrid platforms demonstrates the periodic sequence of strain-induced lateral heterojunctions endowed with asymmetric Schottky vs Ohmic response at the alternating Au-$MoS_2$ nanostripe arrays.

## 2. Results and Discussion

By exploiting the unique capability of thermal Scanning Probe Lithography (t-SPL[46] NanoFrazor instrument) to 'write' high-resolution 3D nanostructures onto thermally labile materials with a hot nanoprobe, we have obtained deterministic grayscale nanotemplates (sketch in **Figure 1a**) which have been exploited as strain engineering platforms for 2D TMD semiconductors.

To achieve high resolution writing capabilities onto optically transparent substrates, we exploit glass substrates coated with a thin indium tin oxide (ITO) film, acting as a counter electrode for the nanolithography process (see methods for details) allowing the precise control of the contact force between t-SPL nanotip and surface[47]. We perform t-SPL on a 140 nm thick Polyphthalaldehyde (PPA) polymer layer spin-coated onto a transparent and conductive ITO thin film supported by a glass substrate.

We exploited t-SPL nanolithography to develop grayscale nanotemplates characterized by arbitrarily defined 3D shape, as shown in the topography map of **Figure 1b**. Here the 3D view of the design fed to the NanoFrazor software (light blue sketch) is reported together with the 3D topography image acquired in real-time by the t-SPL system during the thermal nanolithography, while **Figure 1c** shows the same top view image. The highly ordered nanoarrays are characterized by a periodicity of 530 nm and homogeneous height of the nanostructures of 60 nm. We exploit the high control of the t-SPL tip to fabricate oriented ridges with deterministic tilt angle of the facets, with lateral spatial resolution in the range of 10 nm and with depth-resolution up to 1 nm[48]. The topography image (Figure 1c) highlights the presence of periodic oriented ridges, as demonstrated by the histogram of the slopes (**Figure 1d**). The latter shows a central maximum at 0° corresponding to the inter-ridges flat regions, and two further maxima at -29° and +37° corresponding to the right and left nanofacets, respectively. These high-resolution grayscale patterns have been produced over uniform areas extending hundreds of $\mu m^2$ (see **Figure S1** in Supporting Information), showing the stability of this t-SPL approach onto transparent dielectric substrates.

These t-SPL grayscale templates have been exploited to perform deterministic transfer of mono- and few-layer $MoS_2$ semiconductors[49] (sketch in **Figure 2a**), as shown in the Atomic Force Microscopy (AFM) image of **Figure 2b**.

The AFM image shows nanowrinkled few-layer $MoS_2$ flakes transferred onto the 3D faceted t-SPL template. A 2D monolayer $MoS_2$ (highlighted by a white dashed line) and a few-layer flake are detected, as confirmed by the Raman spectra of **Figure 2c**. The monolayer conformally follows the faceted periodic topography as shown in more detail in the zoomed inset image of Figure 2b with saturated vertical dynamic.

In parallel the wider few-layer $MoS_2$ flake is characterized by a gradually decreasing grade of conformality from top to bottom until it becomes almost completely free-standing in the lower regions (see AFM line profile in **Figure S2** of Supporting Information).

The monolayer nature of the flakes is confirmed by microRaman spectroscopy, using a Jasco NRS-4100 confocal Raman spectrometer with a 100× objective (NA=0.9). Spectra have been detected by excitation of the 2D layer at 532 nm wavelength (details in the Methods section). Figure 2c compares the Raman spectrum acquired on the few-layer $MoS_2$ flake (red line) and on the strained monolayer $MoS_2$ (blue line) shown in Figure 2b. The spectrum of a flat $MoS_2$ monolayer (green line) transferred on the same substrate in proximity to the faceted pattern is also shown for comparison (see image of the flat flake in Supporting Information **Figure S3**). All the spectra show two maxima that are compatible with the vibrational modes $E^1_{2g}$ and $A_{1g}$ of $MoS_2$ layers. The Raman shift of each maximum has been evaluated by a Lorentzian fit of the spectrum, also calculating the spectral shift $\Delta K$ between the $E^1_{2g}$ and the $A_{1g}$. For the few-layer flake, the spectral shift reads $\Delta K_{few\text{-}layer}=(24.1\pm0.5)\ cm^{-1}$, a value that is compatible with a few-layer $MoS_2$. The spectra acquired on the two monolayers show two maxima at about 387 $cm^{-1}$ and 407 $cm^{-1}$, with calculated spectral shifts reading $\Delta K_{2D\text{-}strained}$=19.1 $cm^{-1}$ and $\Delta K_{2D\text{-}flat}$=19.4 $cm^{-1}$ for the strained and for the flat layer, respectively. These values of Raman shift confirm the monolayer structure of these 2D $MoS_2$ flakes [50]. Additionally, we observe a factor 2 enhancement of the Raman intensity of the strained monolayer with respect to the flat configuration, in accordance with recent results on strained TMDs[51]. This first observation suggests a strain-induced modification of the optoelectronic response of the 2D layer conformal to the nanopattern.

To evaluate the optoelectronic response at the microscale we detect the photoluminescence (PL) spectra (**Figure 2d**) acquired both onto the strained and onto the flat 2D monolayer. We detect an enhancement of the photoluminescent emission of the strained 2D layer with respect to the flat layer of factor 2.9 for A-exciton emission (details in Supporting Information, **Figure S4**). In both the cases a strong exciton emission is detected at 1.89 eV energy due to the direct radiative recombination of A exciton, superimposed to a contribution of the trion-emission at room temperature [52]. A weaker emission signal is detected at 2.03 eV, corresponding to the radiative recombination of B-exciton.

The PL measurements show the capability to modify the photonic response of 2D TMD semiconductor layers by using these periodic faceted nanotemplates. However, to better resolve strain induced effects that are expected to occur at the local scale, we investigate the electronic response of the rippled 2D semiconductor layers via Kelvin Probe Force Microscopy (KPFM). This probe-based technique indeed allows the imaging of the surface potential with high spatial resolution [53] We perform KPFM imaging of the wrinkled 2D $MoS_2$ in single-pass mode, that guarantees the co-localized acquisition of topography and surface potential signal by using the Nano-Observer instrument by CSI.

**Figure 3a** and **Figure 3b** show the topography and the co-localized Contact Potential Difference (CPD) in Kelvin probe microscopy of the nanowrinkled $MoS_2$ layers. The detected CPD signal is defined as follows:

$$CPD = \frac{\phi_{tip} - \phi_{surf}}{e}$$

Where $\Phi_{tip}$ is the tip work function and $\phi_{surf}$ is the work function of the surface, $e$ is the elementary charge. A strong CPD contrast of about 100 mV is detected corresponding to $MoS_2$ monolayer with respect to the supporting template (Figure 3b), despite its topography contrast is very weak (white dashed line in Figure 3a). In parallel, a strong *CPD* contrast is also detected corresponding to the few-layer flake.

The work function difference between monolayer and few-layer $MoS_2$ has been evaluated by considering the flat inter-ripple regions and extracting the histogram of the detected CPD (**Figure 3c**), as described in detail in **Figure S5** of Supporting Information. A two-component gaussian fit is calculated in correspondence to the lower and a higher CPD distributions that correspond to the few layer (red line) and to the monolayer $MoS_2$ (blue line), respectively. The gap between the two distributions reads $\Delta_{CPD}$= 240 mV, that corresponds to the work function difference between flat monolayer and few layer $MoS_2$ [54]. Remarkably, in the CPD image of Figure 3b, the signal of both the monolayer and the few-layer $MoS_2$ also shows a periodical modulation in register with the underlying nanopattern. Asymmetric CPD dips are detected onto the left facet of each nanoridge, as shown by the line-profiles of **Figure 3d** and **Figure 3e** (blue lines), respectively corresponding to the strained monolayer and few-layer $MoS_2$. The periodic surface potential dips read about 130 mV for monolayer, and 150 mV for few-layer $MoS_2$, and are detected with high spatial resolution at the level of few tens of nanometers. These characteristic periodic features are due to the conformal adhesion of the 2D layer onto the template and homogeneously run along the whole rippled 2D layer structure in register with the pattern (darker regions in Figure 2b, further CPD line profiles in **Figure S6** of Supporting Information). Indeed, for few-layer $MoS_2$, the modulation of the CPD is reduced near the edges of the flake, where conformality is not fully preserved, as shown in **Figure S7**.

This periodic and asymmetric modulation of the surface potential can be attributed to template driven strain, rising during transfer of the 2D TMD layers from the soft Polydimethylsiloxane (PDMS) elastomeric stamp. Indeed, the release of the 2D materials onto the periodic templates occurs along a direction perpendicular to the templates ridges thus inducing an asymmetric field strain in the material, as shown by the sketches in **Figure 3f**. Periodic CPD minima shown in Figure 3d and Figure 3e can be attributed to a pattern-modulated tensile strain homogeneously localized along the left-side of each nanoridge. The electrical work function increase (i.e. CPD decrease) detected along the left side nanofacets can be thus associated to a tensile strain accumulation as also shown in recent simulation studies[25].

In our experiment the difference between the elastic properties of the PDMS stamp supporting the flat 2D TMDs, and the PPA template allows the conformal transfer of the 2D layers onto the faceted nanopatterns. The Young modulus of the PDMS and of the PPA reads about $8\ 10^{-4}$ GPa and 0.3 GPa respectively[55,56], thus suggesting that the PDMS stamp strongly deforms onto the nanostructured PPA template characterized by higher stiffness. As shown in the sketches of Figure 3f, and more in detail in **Figure S8** of Supporting Information, the 2D layer first lies onto the right-side facets without deformation, then it follows the PDMS conformal stretching onto the left-side facets of the nanotemplate. After the complete conformal adhesion of the flake, the PDMS is released leaving a non-homogeneous tensile strain into the 2D layer that is periodically localized along the left-side facets.

This behavior demonstrates the capability to tailor the electronic properties of 2D semiconductor layers at the nanoscale level in a controlled way by controlled transfer of 2D layers onto faceted nanotemplates fabricated by grayscale t-SPL.

These results offer the possibility to devise deterministic nanoarrays of lateral 2D heterojunctions based on strained 2D TMDs. To provide a proof-of-concept configuration we engineer hybrid 2D-metallic nanoarrays with the aim to contact specific portions of 2D TMD layers characterized by different degree of strain. To achieve this goal we exploit the opportunity provided by the grayscale faceted nanopatterns that allow the maskless confinement of periodic metal nanowires (NWs) by glancing angle Au evaporation perpendicular to the ripple ridges direction, at 80° with respect to the surface normal (sketch in **Figure 4a**). The effective confinement of gold NWs is confirmed by high resolution AFM imaging and SEM characterization shown in Supporting Information, **Figure S9**. The conformal adhesion of the 2D TMD layers onto the template allows us to decorate the 2D TMD semiconductor layers with the periodic Au NWs that act as local nanocontacts for the 2D material, as shown by the AFM topography of Figure 4a. This approach brings a two-fold advantage since i) the Au NWs acting as conducting nanoelectrodes are deposited directly on the 2D materials without need for further lithographic steps, thus avoiding unwanted doping and/or damaging of the fragile 2D material. Additionally, ii) the NWs electrodes address portions of the 2D material which are periodically strained at the nanoscale showing template-modulated electronic surface properties.

To investigate the electrical behavior of these hybrid 2D TMDs-metallic layers, we perform local KPFM measurements. The KPFM map of **Figure 4b** shows a strong CPD contrast with periodic behavior colocalized with the topography image of Figure 4a. Tilted Au nanostripes (bright regions) are detected with high spatial resolution in both the topography and in the CPD map (Figure 4b). These NWs are characterized by typical width of about 300 nm and are decorated at their edge by chains of Au nanoclusters that are resolved in both the topography and the CPD map (inverted color scale). The CPD is lower onto the Au NWs and reads about 170 mV, while it reads about 315 mV onto the bare PPA template. The hybrid 2D-metallic layer is detected at the center of the image (dashed white line in the topography) inducing further modulation of the detected CPD signal. In this configuration, a proof-of-concept device based on periodic Au-$MoS_2$-Au lateral heterojunctions is achieved. The colocalized CPD map (Figure 4b) and a specific line profile (**Figure 4c**) show a periodic modulation of the electric response at the nanoscale, in register with the template. A lower CPD value reading about 200 mV is detected at the left Au-$MoS_2$ heterojunction (green dashed lines in Figure 4b and Figure 4c) while a higher CPD value reading about 290 mV is detected at the right $MoS_2$-Au heterojunction. This asymmetric electrical behavior is homogeneously detected onto the whole 2D-metallic hybrid structure confirming the capability to tailor at will the electronic properties by the devised strain engineering approach.

To understand this effect in Figure 4c we compare the CPD profile of the Au-$MoS_2$ heterojunctions (continuous blue line) with the corresponding CPD profile detected onto the bare strained $MoS_2$ (dashed blue line extracted from Figure 3d). The two detected CPD profiles have been normalized to the work function of the tip ($\Phi_{tip}$) corresponding to the CPD map of Figure 4b (details in Supporting Information **Figure S10**). The value of $MoS_2$ surface potential at the left heterojunction (dashed green lines in Figure 4 c) does not change due to Au NWs deposition, reading a constant value of about 200 mV. This suggests the presence of an Ohmic junction given by the local match of the electronic work functions at the Au-$MoS_2$ lateral interface. At the right heterojunction (dashed blue line in Figure 4c) we observe a strong increase of the surface potential that can be attributed to a built-in Schottky barrier at the strained $MoS_2$-Au lateral interface. This effect can be explained in terms of the strain induced modulation of the Fermi Level Pinning and of the Charge Neutrality Level, as recently reported [57,58]. These results show the possibility to tune the electronic surface properties of 2D TMDs at the local scale by simply tailoring their shape in a strain engineering configuration enabled by thermal nanolithography. This offers the unique opportunity to create 2D nanoelectronic and nanophotonic devices endowed with asymmetric local lateral junctions, featuring optimized carriers' extraction with strong impact on their performances.

## 3. Conclusion

We developed a novel strain engineering approach for 2D TMD semiconductor layers based on the grayscale thermal- Scanning Probe Lithography that allows to tailor the electronic properties of 2D materials at the nanoscale. These locally strained 2D layers enable the fabrication of asymmetric lateral heterojunctions by simple maskless deposition of Au nanocontacts.

Periodic grayscale nanotemplates are fabricated by t-SPL achieving deterministic faceted nanoridges with nanoscale lateral and vertical spatial resolution. These peculiar 3D templates are exploited for the transfer of 2D $MoS_2$ semiconductor layers that conformally follows the underlying faceted templates. This way the 2D TMDs monolayer shows an amplification of the photoluminescence signal detected at the microscopic scale, suggesting strain accumulation in the 2D material. To locally resolve this effect, we perform KPFM imaging of the surface, detecting a periodic modulation of the electrical work function which is asymmetric with respect to each faceted nanoridge. This local modulation of the electronic properties allows to develop lateral heterojunction nanoarrays based on hybrid Au-$MoS_2$ layers. Thanks to the faceted morphology these hybrid nanoarrays are achieved by maskless glancing angle metal deposition onto the faceted templated without affecting the quality of the 2D material with further lithographic steps. This platform shows periodic lateral Au-$MoS_2$ heterojunctions driven by the template, characterized by asymmetric Ohmic vs Schottky electronic behavior. These strained 2D TMD semiconductor platforms thus allow to engineer carrier extraction functionalities in novel ultra-thin lateral junctions devices, with impact in many fields ranging from nanoelectronics and nanophotonics to energy conversion and sensing.

## 4. Experimental Methods

*Sample preparation and thermal-Scanning Probe Lithography nanofabrication:*

Glass substrates (15x20 $mm^2$ from Ossila) coated with 100 nm thick ITO film has been exploited for sample fabrication. Since grayscale lithography by t-SPL needs for a conductive counter-electrode, a silver paste electrode deposited onto the conductive film has been exploited.

A PPA film of 160 nm thickness is deposited on the sample by spin coating of a 8% PPA:Anisole solution prepared with Phoenix 81 powder by AllResist. Spin coating is followed by annealing at 110°C. Reported thickness values are observed by AFM imaging of the film across a scratch on the film exposing the underlying substrate with a sharp edge.

t-SPL nanolithography has been performed by using NanoFrazor Scholar instrument from Heidelberg Instruments Nano, under temperature of the tip in the range of 800-1100°C. The conductive sample-holder of the NanoFrazor, acting as counter-electrode for the nanotip, has been electrically connected to the ITO coated substrates.

*Exfoliation and Deterministic transfer of 2D TMD layers onto t-SPL nanopatterns:*

Exfoliation of 2D $MoS_2$ layers was performed by both scotch tape (Nitto SPV224 PVC tape) and PDMS from a bulk $MoS_2$ crystal (commercial material by SPI). The scotch tape was used for the initial reduction of the thickness of macroscopic single crystals down to sub-micrometric range. Then, the PDMS was exploited to obtain atomically thin flakes. In a final step, the 2D $MoS_2$ layers have been transferred onto the t-SPL nanopattern using a deterministic viscoelastic transfer set-up (**Figure S11**).

*Kelvin Probe and Atomic Force Microscopy characterization:*

Kelvin Probe Force Microscopy analysis is performed in ambient conditions by using CSI Nano Observer instrument operating in KPFM-Single Pass mode. Electrical contact is formed between the grounded sample holder and the ITO surface by use of a copper wire and silver paste. For the characterization, we used Pt coated AFM tips (ANSCM-PT) from AppNano. Atomic Force Microscopy (AFM) characterization of the sample have been performed by Neaspec instrument (attocube).

# Figures

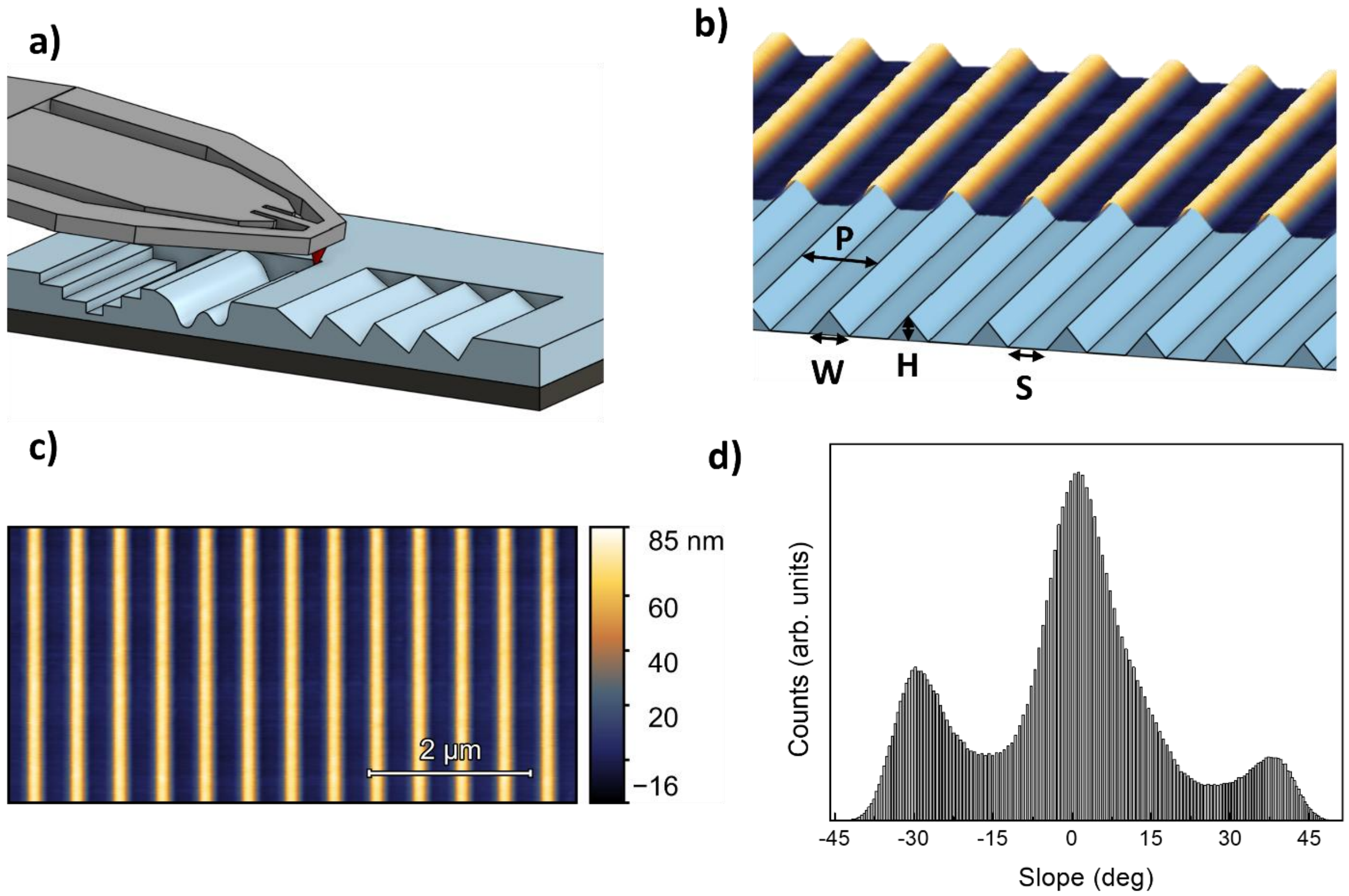

**Figure 1.** a) Sketch of the grayscale t-SPL process on a PPA layer. b) Three-dimensional view of the faceted structure as designed (bottom) and overlayed measured AFM topography after patterning (top). c) Top view of AFM topography after patterning. d) Slopes distribution extracted from c) as the first derivative of the line profiles.

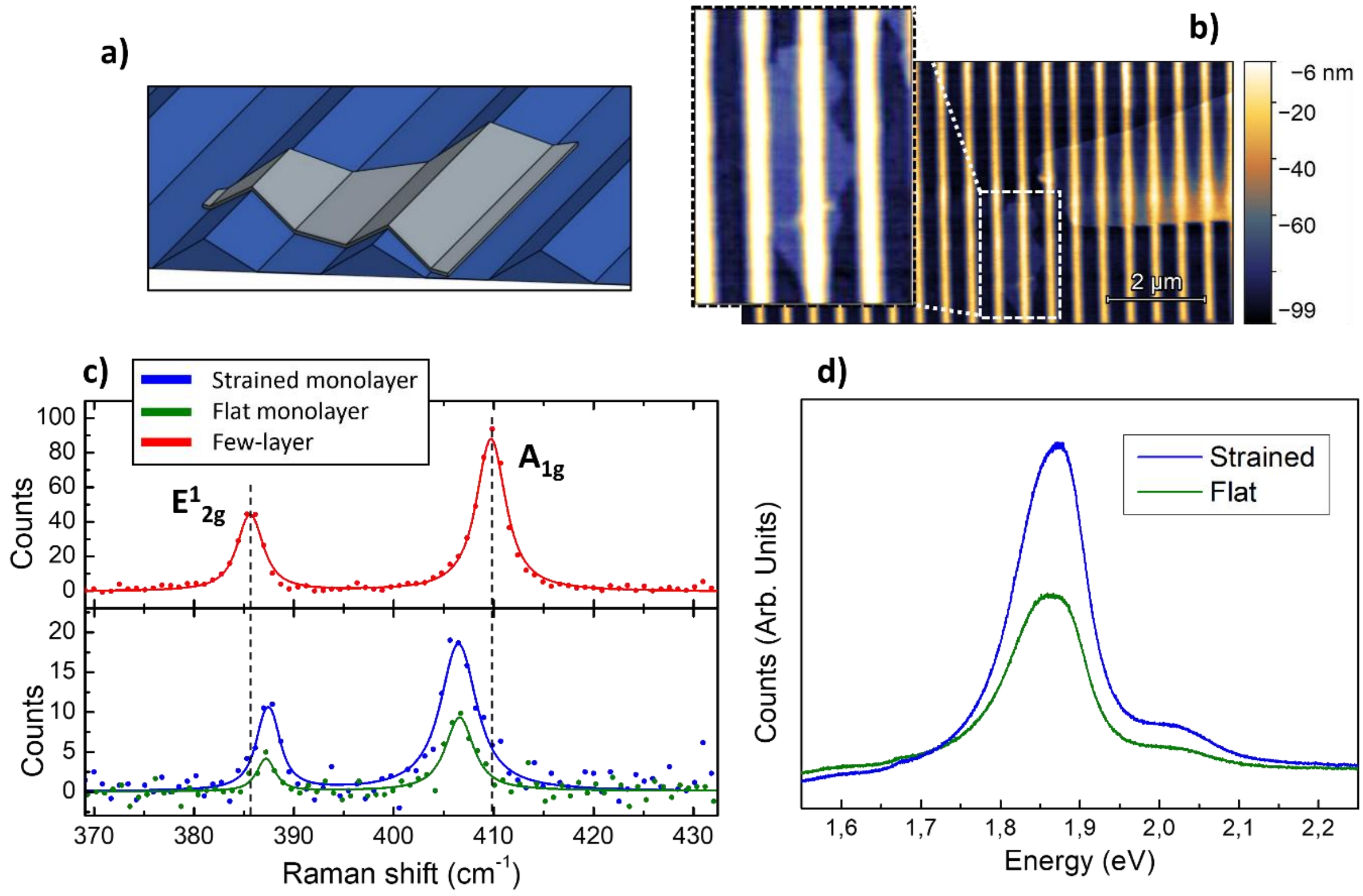

**Figure 2.** a) Sketch of a 2D $MoS_2$ layer conformal to the underlying template. b) AFM image few-layer $MoS_2$ conformally transferred onto the grayscale nanotemplate. A zoom on the 2D monolayer is shown in inset with saturated vertical dynamic. c) Raman spectra of the few-layer $MoS_2$ flakes: the strained few-layer (red line), the strained monolayer (blue line) and a flat monolayer lying on the same substrate (green line) are shown. d) Photoluminescence spectra of the strained (blue line) and flat (green line) $MoS_2$ monolayers.

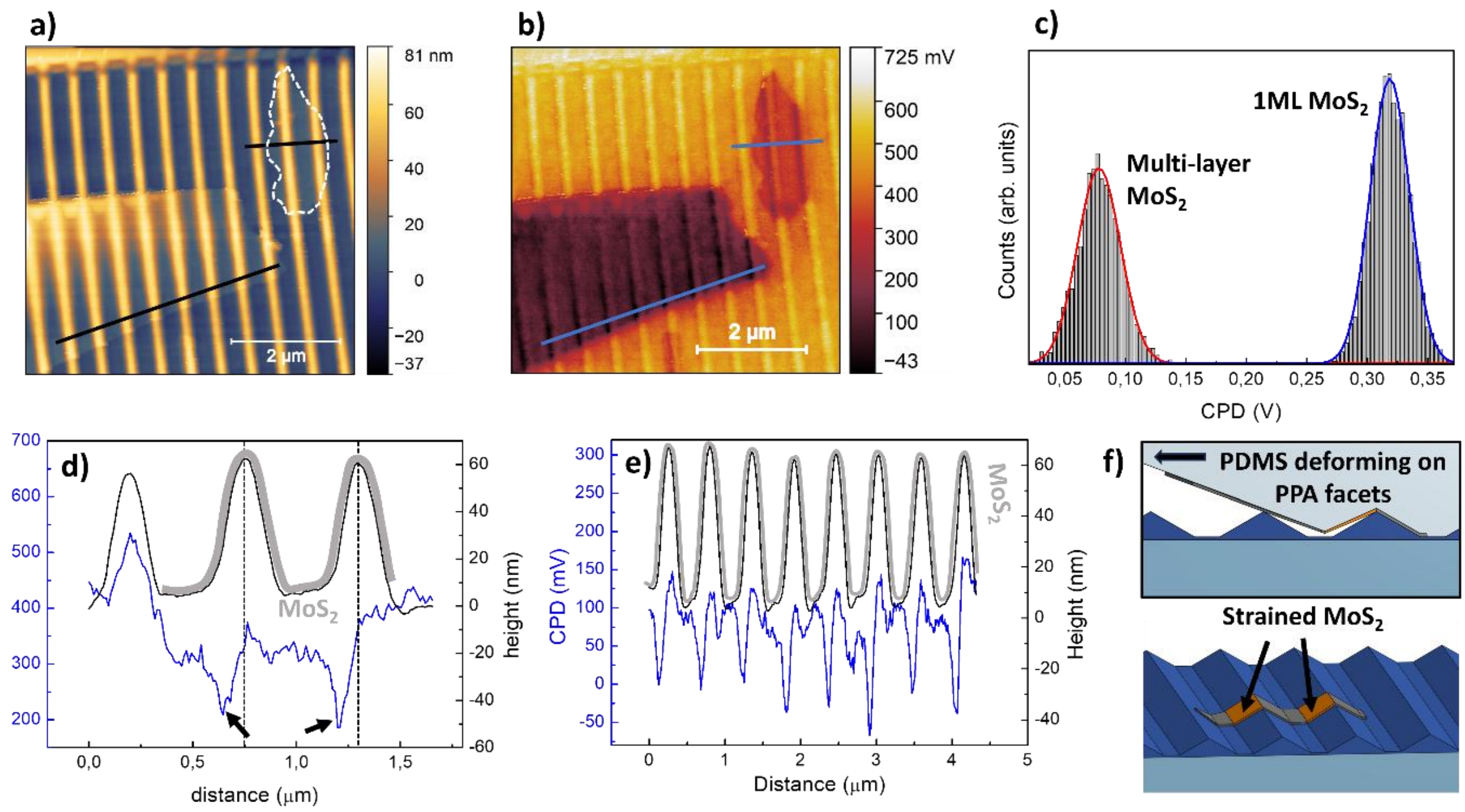

**Figure 3.** a-b) Co-localized AFM and CPD image of the transferred 2D monolayer and multilayer $MoS_2$ flakes detected by single-pass KPFM. c) Histogram of CPD data extrapolated from selected flat regions of mono- and multi-layer $MoS_2$. d-e) Co-localized topography (black lines) and CPD signal (blue lines) corresponding to mono- and multi-layer $MoS_2$, respectively, extracted from the line profiles highlighted in a) and b). The regions where a $MoS_2$ layer is present are depicted with thick grey lines. f) Sketch of the strain accumulation effect during PDMS-mediated transfer of the 2D materials onto the nanotemplate, and periodic modulation of the strain in the nanowrinkled 2D material.

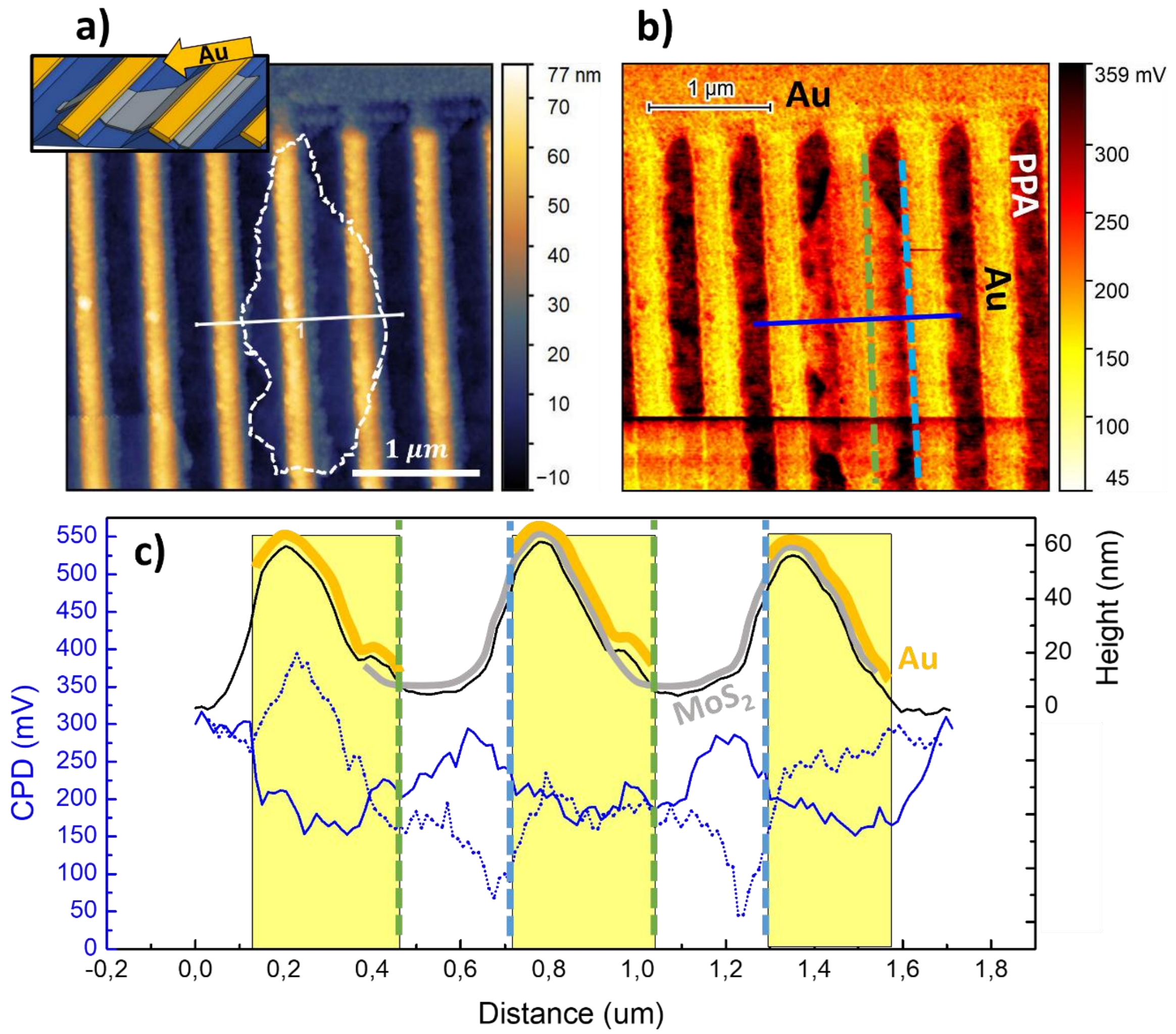

**Figure 4.** a-b) Topography, sketch and CPD map of hybrid Au-$MoS_2$ nanoarrays. c) Co-localized line profiles extracted from maps in a) and b). Regions corresponding to $MoS_2$ and Au-$MoS_2$ nanosystems are highlighted with grey and yellow lines, while green and light blue dashed lines in b, c) correspond to left and right Au-$MoS_2$ lateral heterojunction.

# Supporting Information

## Figure S1: Template nanopatterning on areas exceeding hundreds of μm$^2$:

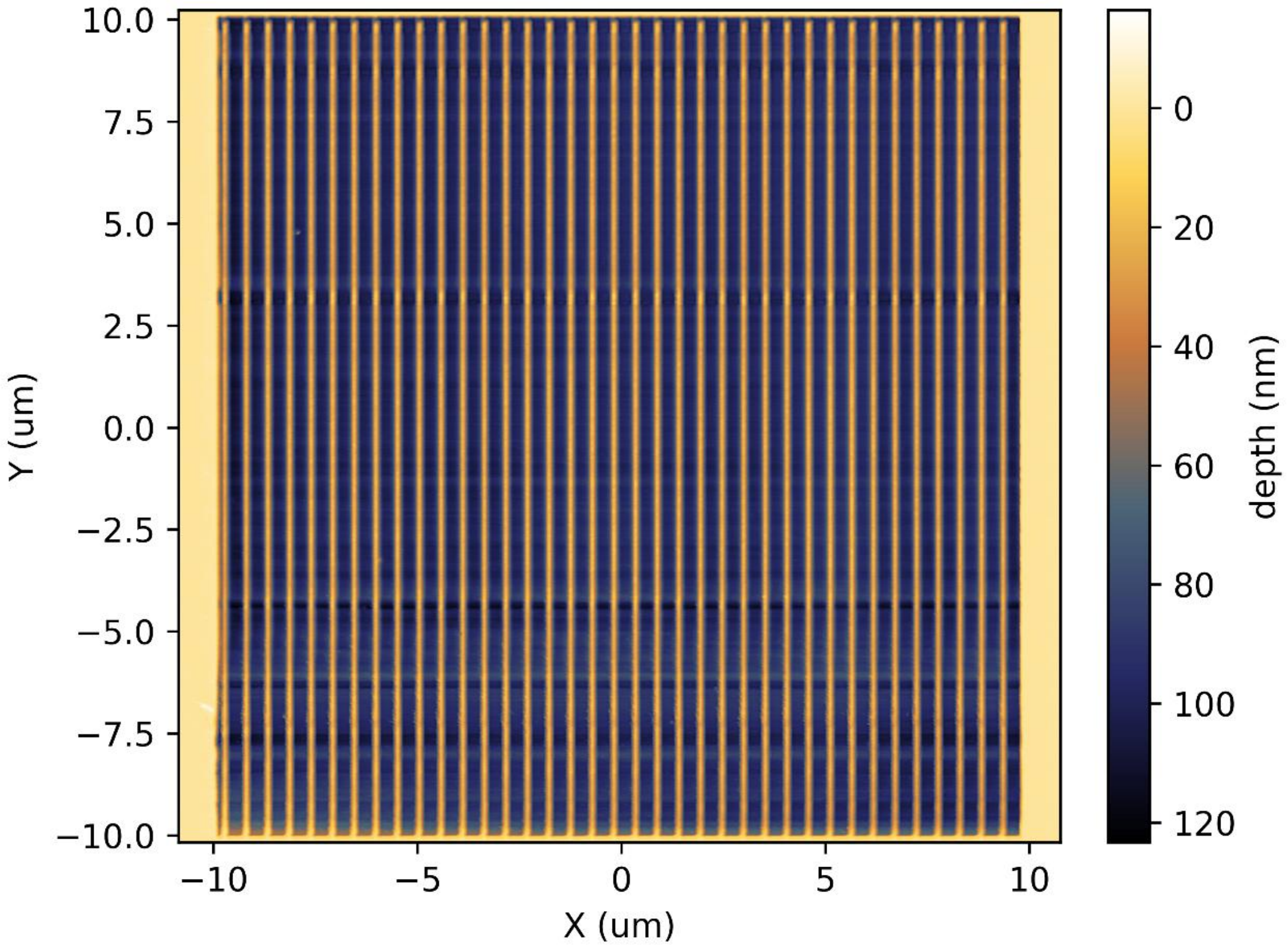

*Figure S1: AFM topography of one of the employed nanotemplates, showing the stability of the patterning technique on an area of around 400 μm$^2$*

## Figure S2: Conformality of $MoS_2$ flakes to the nanotemplate

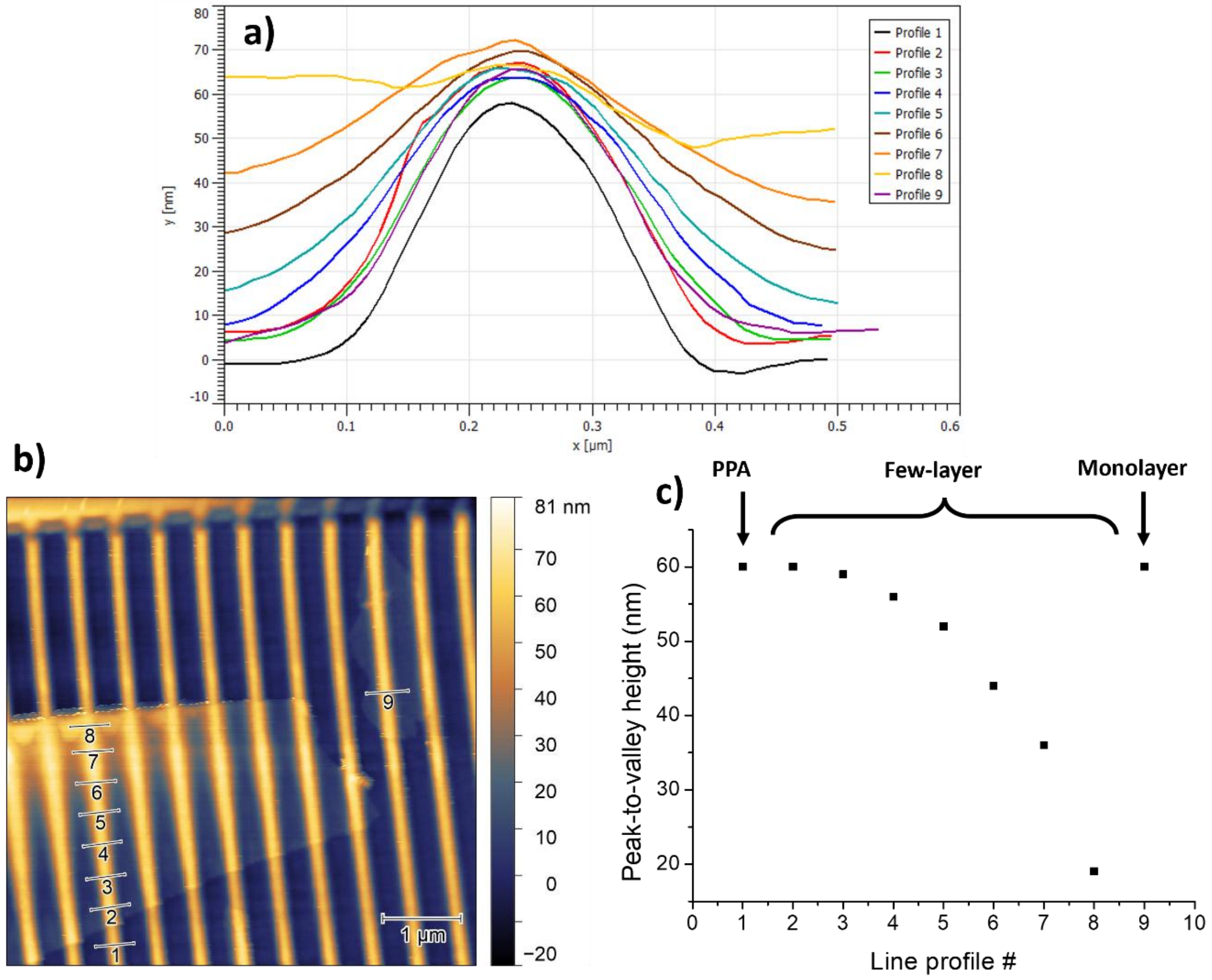

*Figure S2: a) Topography line profiles extracted from the AFM image of Figure 3a in the manuscript. Locations of the extracted line profiles respect to the topography AFM map are reported in b). c) Peak to valley heights as measured from line profiles in a).*

## Figure S3: Flat $MoS_2$ monolayer on unpatterned PPA

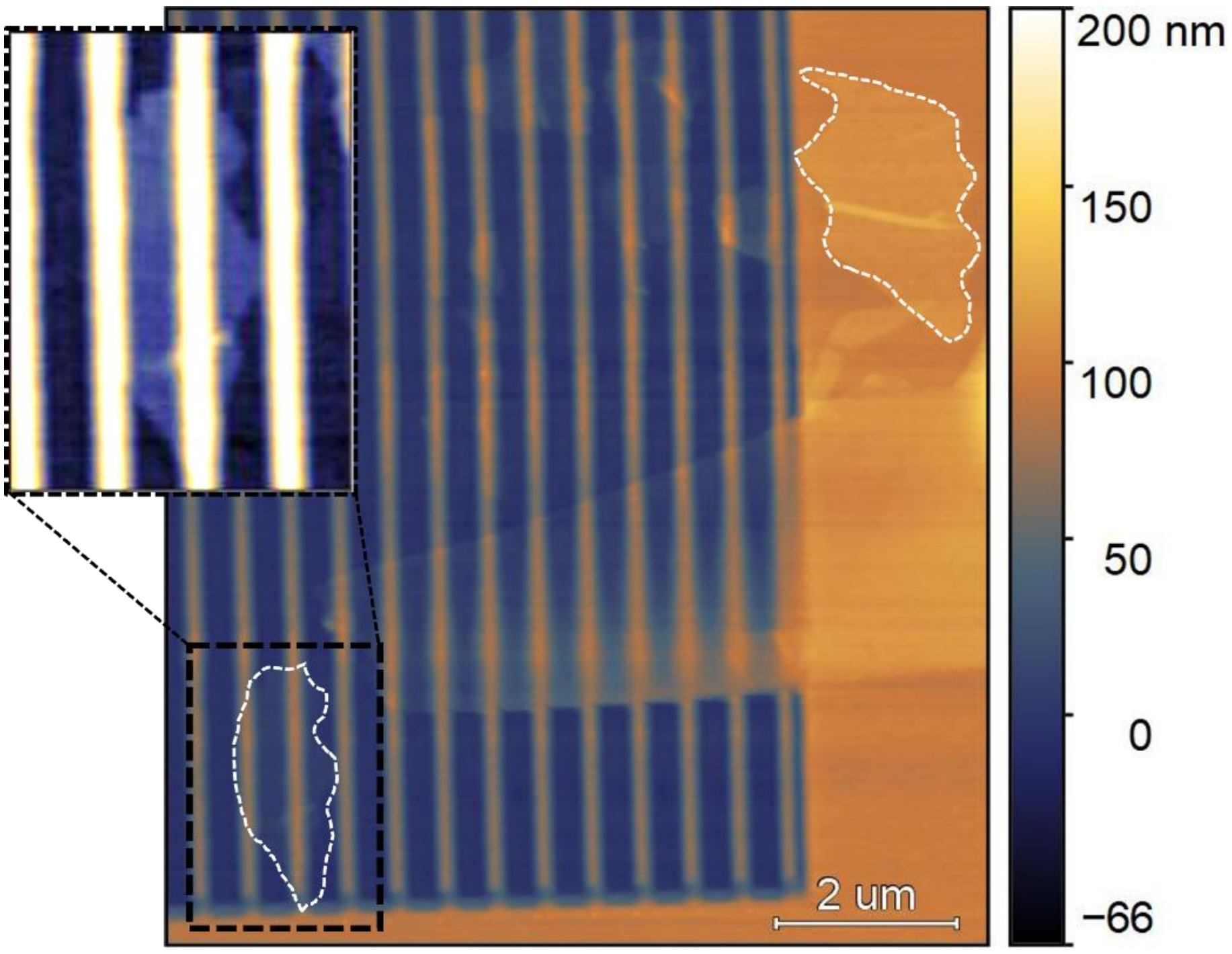

*Figure S3: Extended view of the AFM map reported in Figure 2b in the manuscript, showing the presence of a transferred flat monolayer on unpatterned PPA. Raman and PL spectra of this flake are reported in Figure 2c-d in the manuscript.*

# Figure S4: Fit of PL spectra

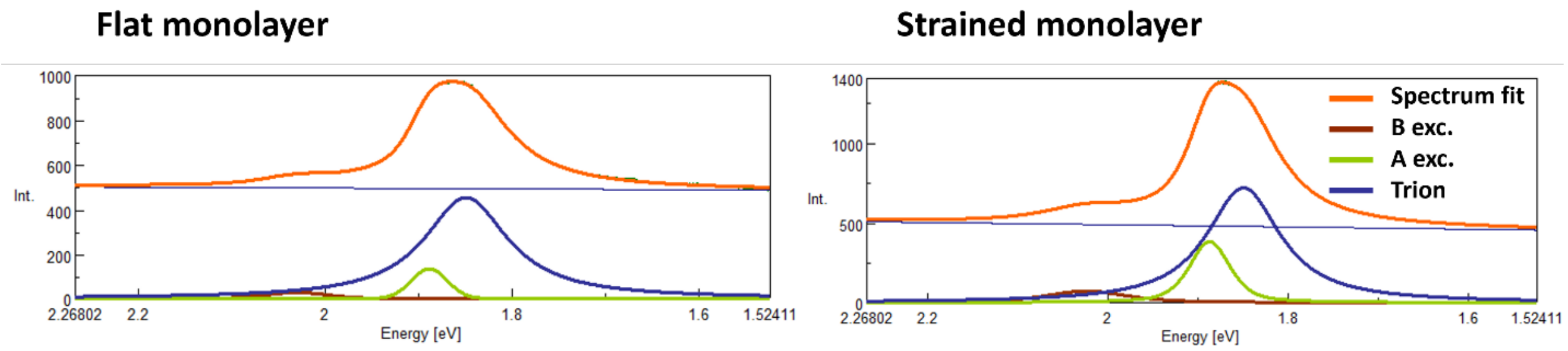

*Figure S4: Fitting results of the PL spectra shown in Figure 2d. The orange line represents the total spectrum fit, while the brown, green, and blue lines correspond to the three Gaussian-Lorentzian components for the A and B excitons and the trion emission. For the two spectra, acquired under identical experimental conditions, the A-exciton emission peak reaches heights of 133 and 380 counts for the flat and strained conditions, respectively, resulting in an enhancement factor of 2.9.*

**Figure S5: Histogram of CPD values for flat mono- and few- layer $MoS_2$**

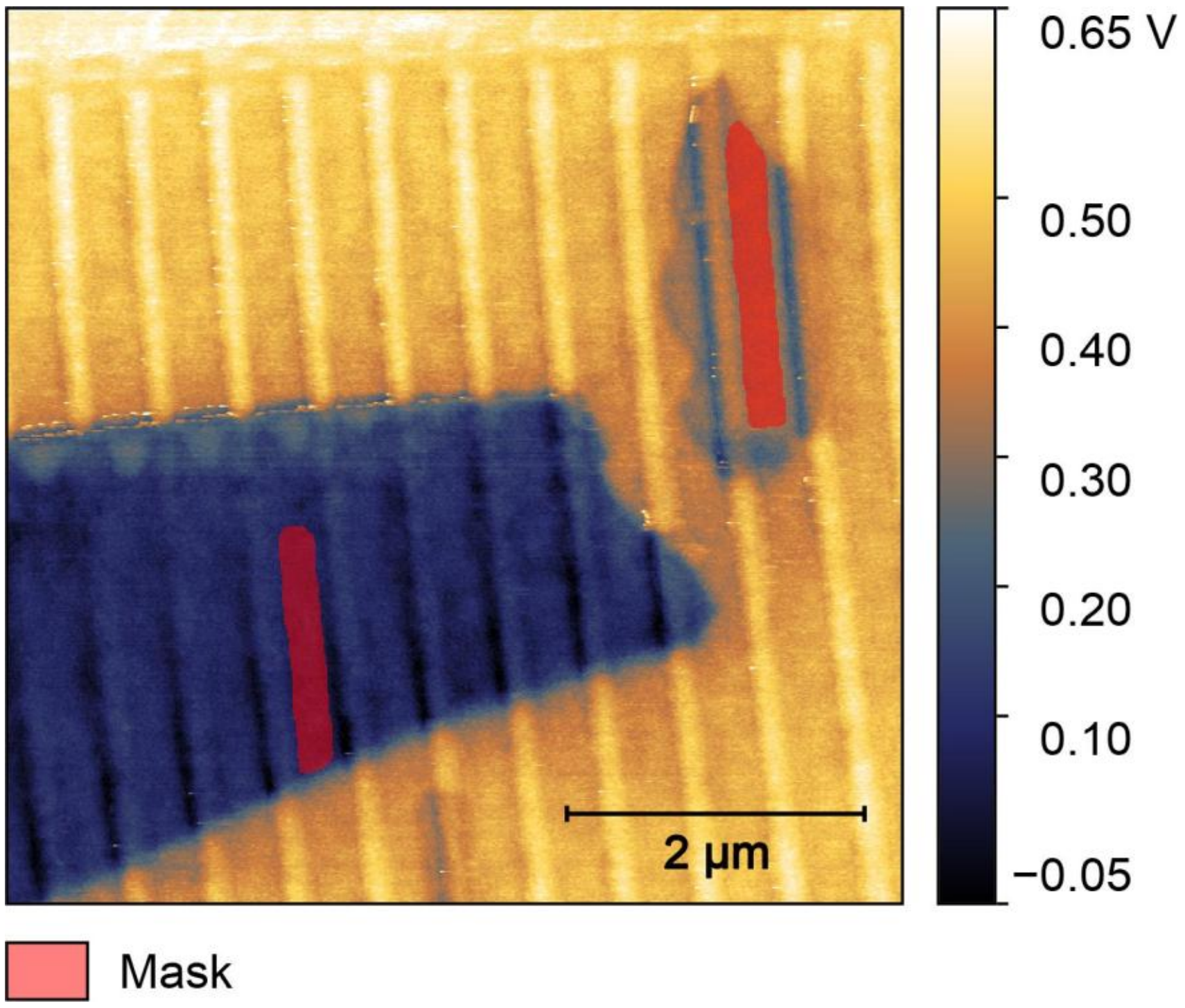

*Figure S5: This CPD map (already shown in Figure 3b in the manuscript with different color scale) shows with a red overlayed mask the portions considered to plot the CPD histogram reported in Figure 3c. Only flat portions of both flakes have been selected to gather CPD informations on flat mono- and few-layer MoS2 in the same KPFM map.*

## Figure S6: CPD profiles

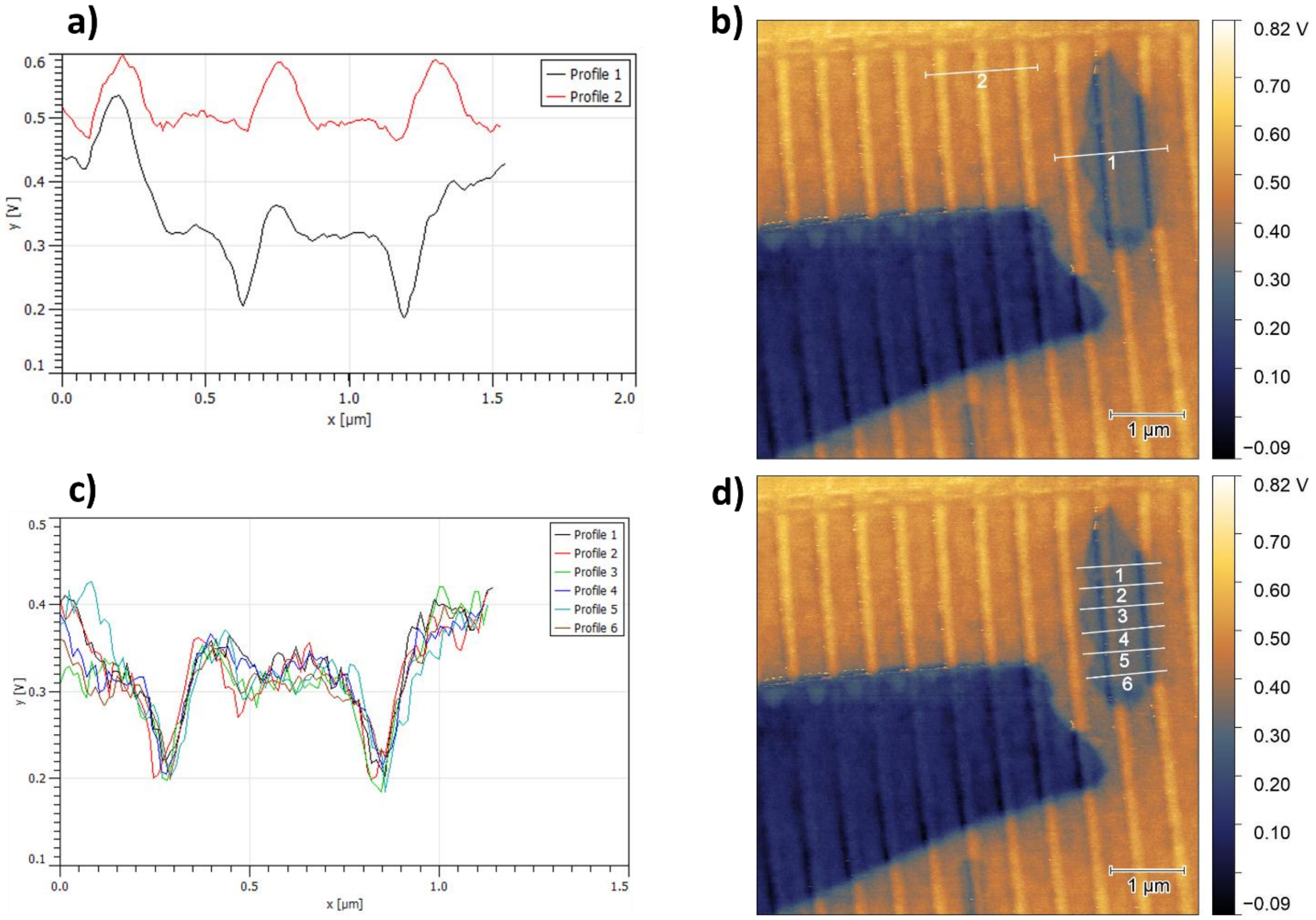

*Figure S6: a-b) CPD profiles extracted from bare patterned PPA and strained conformal monolayer, showing the difference in the measured surface potential. The CPD asymmetry respect to the single ridges is only visible on the strained MoS2 and absent on bare patterned PPA. c-d) Multiple profiles extracted from the strained monolayer, showing that the CPD periodic features discussed in the manuscript run along the whole rippled 2D layer structure.*

## Figure S7: CPD modulation on regions with different degrees of conformality

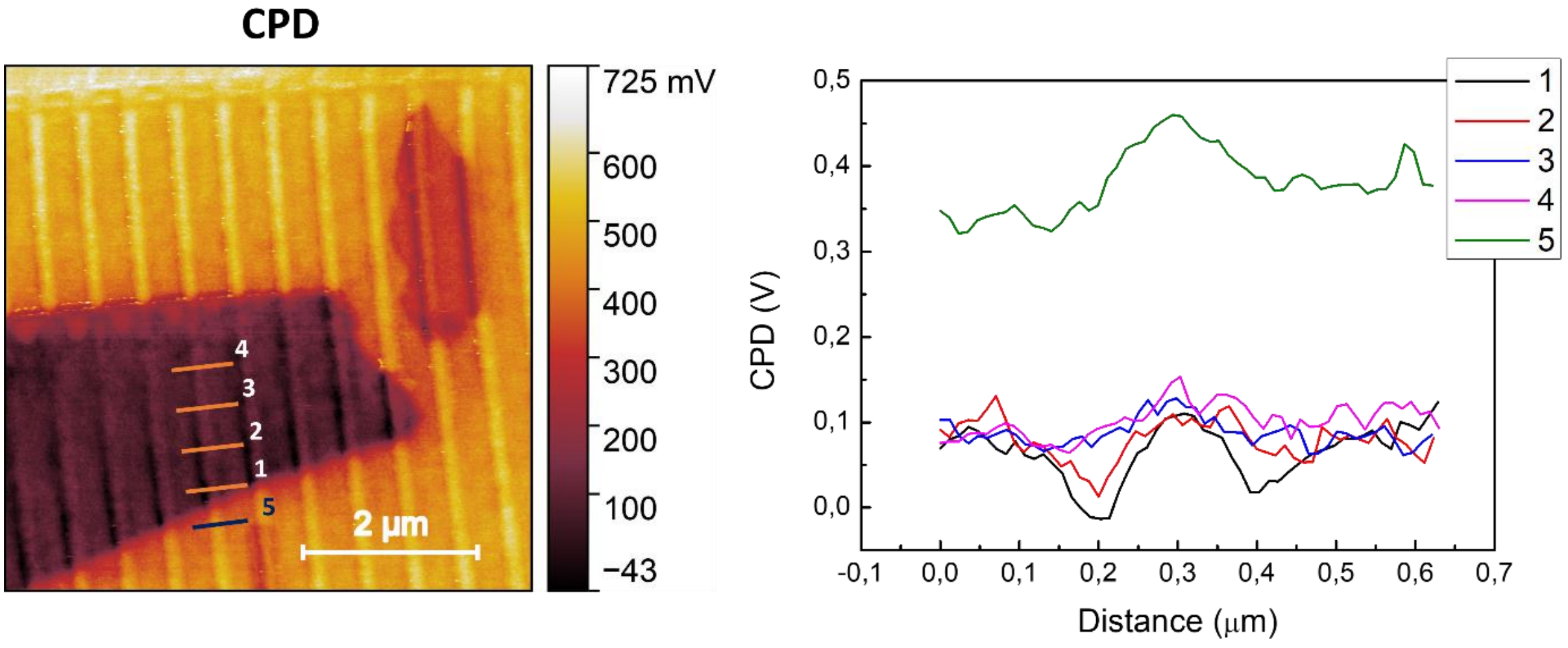

*Figure S7: a-b) CPD map and corresponding CPD line profiles extracted from regions of multi-layer $MoS_2$ with different degree of conformality. A strong CPD modulation is detected in conformal regions (profiles 1, 2) while weaker features are detected in the other cases.*

## Figure S8: Direction of transfer respect to pattern ridges

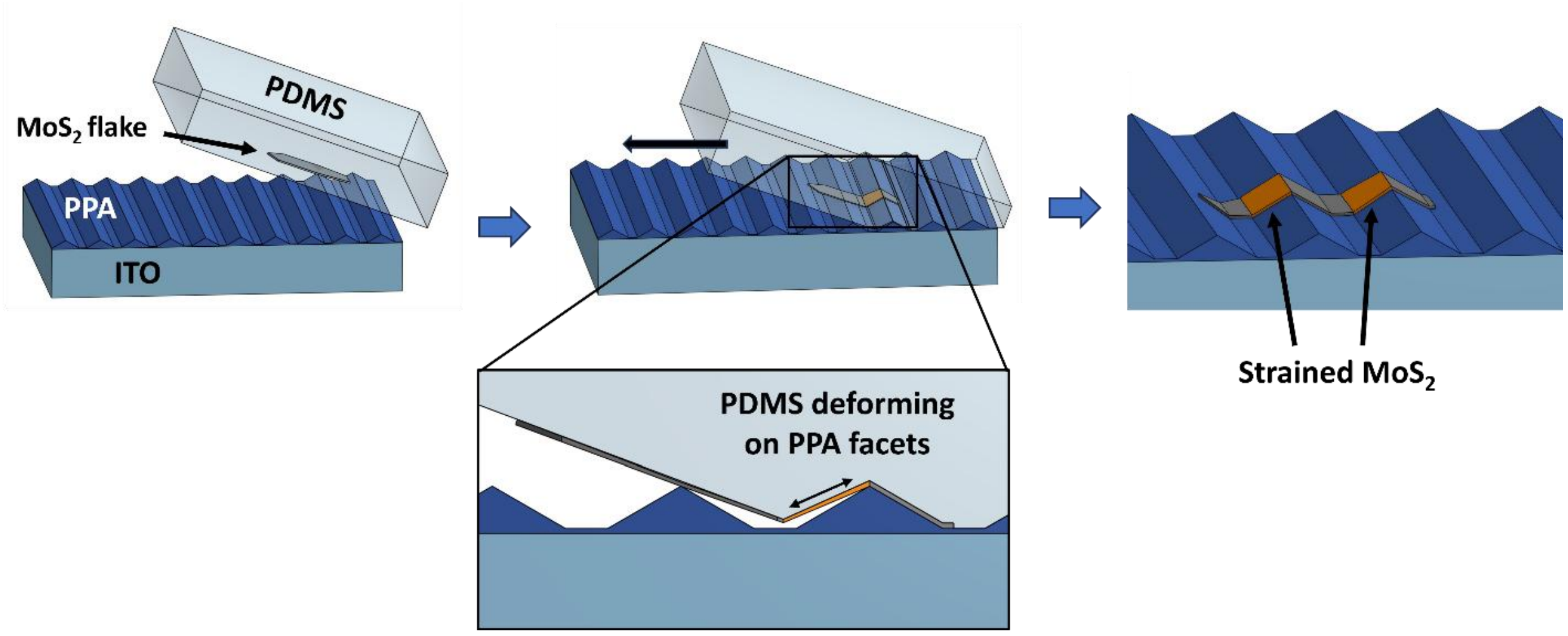

*Figure S8: Schematic view of the directional transfer of MoS2 flakes along the faceted patterns, showing the mechanism behind asymmetric localized straining of the MoS2 on the left facets*

# Figure S9: Characterization of Au Nanowires

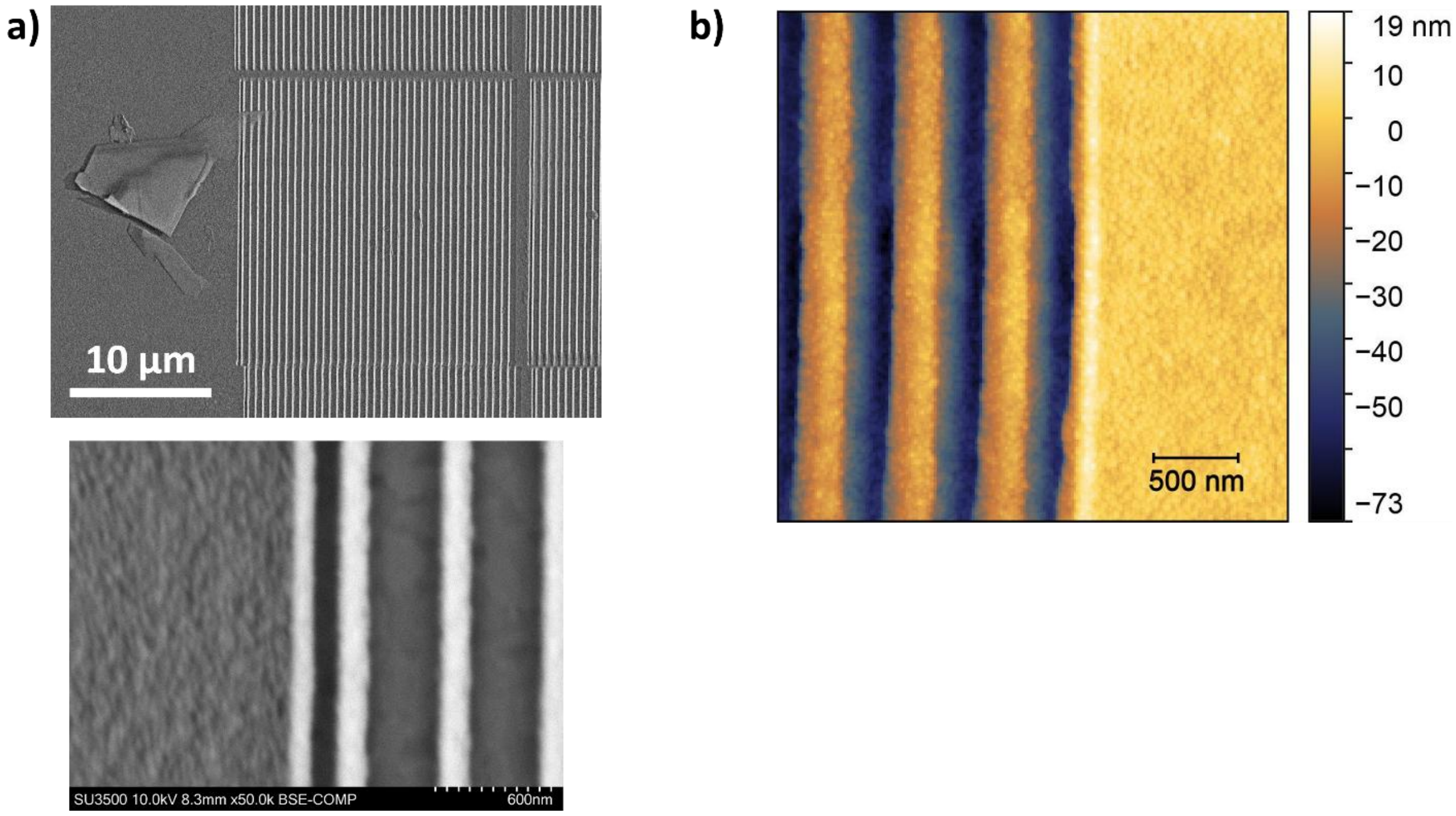

*Figure S9: a) SEM images (BSE signal) acquired on the faceted nanopattern after glancing angle Au thermal evaporation. Au nanowires appear as bright high z-contrast lines, while Au deposited on flat PPA regions show lower contrast. Dark areas between Au NWs represent bare PPA regions. Inset: zoomed SEM detail at the edge of the pattern. b) Detail of a similar pattern measured by high-resolution AFM after glancing angle Au evaporation. The variation in roughness between Au on NWs and on flat PPA is due to the change in local deposition angle and thicknesses, with consequent different percolating behavior*

## Figure S10: Normalization of CPD profiles

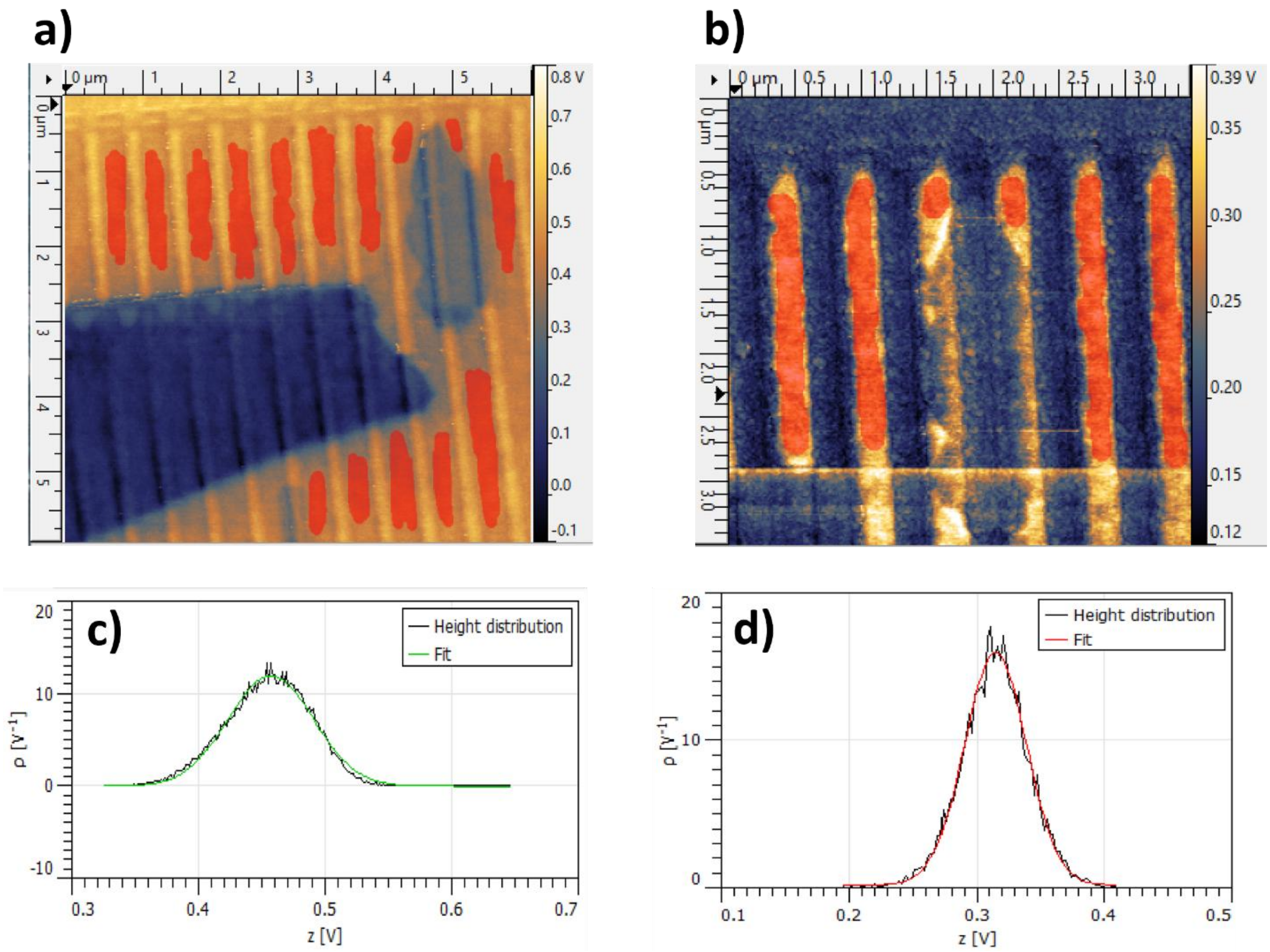

*Figure S10: a-b) CPD maps reported in Figure 3b and 4b in the manuscript, with overlayed red mask. The masks here isolate the areas where bare PPA is exposed, not covered by MoS2 flakes or Gold. c-d) Histogram of CPD values extracted from the areas selected by the masking shown in a-b. Solid lines show Gaussian fits to the two histograms. Centers of the two Gaussian lie respectively at $CPD_{before}$=456 mV and $CPD_{after}$=315 mV*

## Figure S11: Deterministic transfer of flakes on PPA nanopatterns

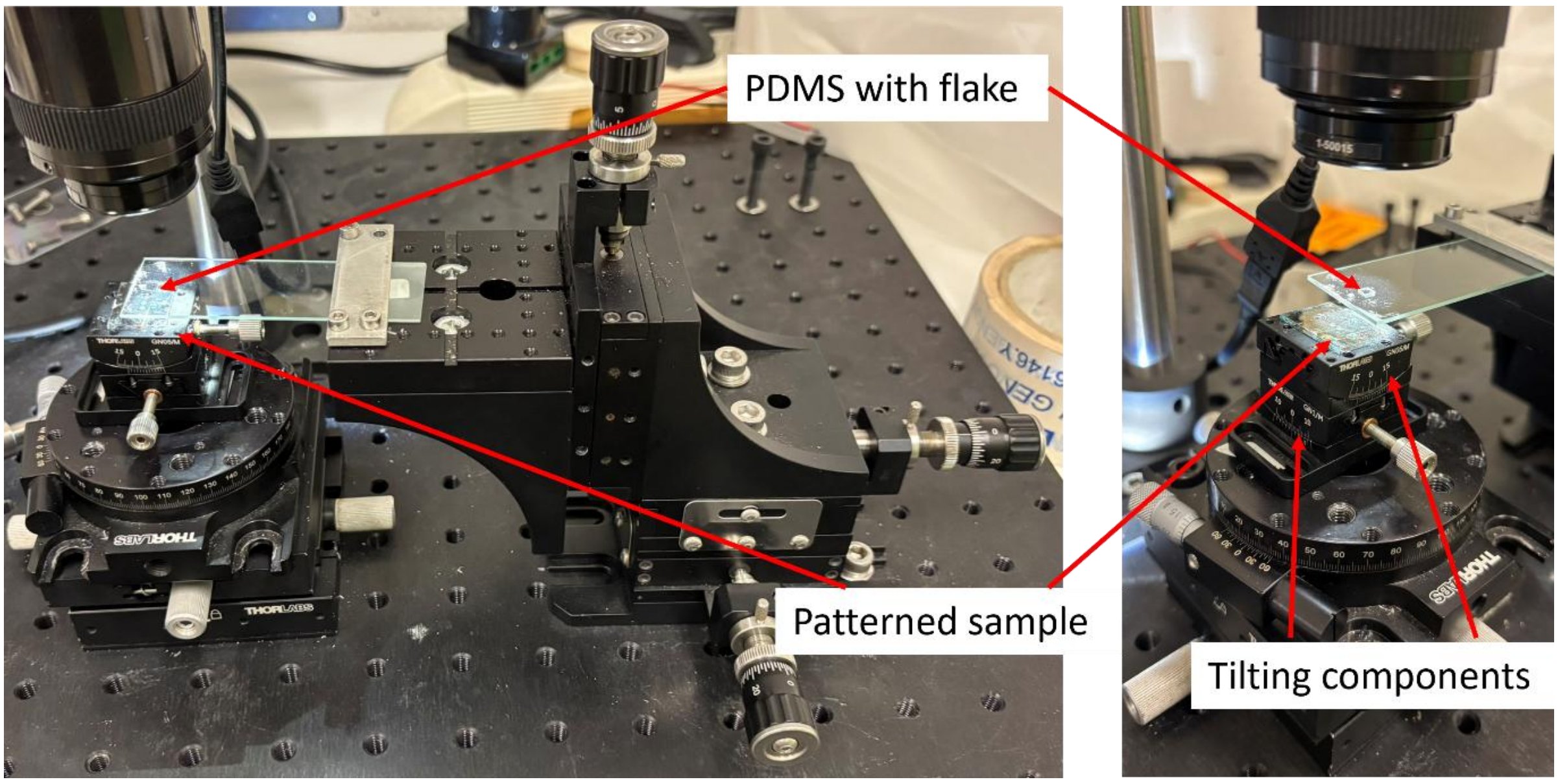

*Figure S11 Picture of the custom optical microscopy setup employed for deterministic transfer of 2D layers.*

The deterministic viscoelastic transfer set-up used for stamping the 2D $MoS_2$ layers onto the t-SPL nanopattern is presented in Figure SI 10. The patterned sample is placed onto a XY translation stage with tilting and rotating platform, while the microscope slide with the PDMS sheet is moved by a XYZ translation stage with standard micrometers. Firstly, the PDMS approaches the substrate face down and the desired flake is centered in the landing position. Then, after the contact is slowly established, finely controlling the contact front, the PDMS sheet raises, releasing the flake onto the patterned sample. The entire process can be controlled by a camera connected to a microscope column[49].